\documentclass[a4paper]{article}
\usepackage{graphicx}

\begin{document}

\begin{center}

{\bf \LARGE Weinberg and few-nucleon forces}

\vspace{1cm}

{\bf \large U van Kolck}

\vspace{0.3cm}

Universit\'e Paris-Saclay, CNRS/IN2P3, IJCLab, 91405 Orsay, France

Department of Physics, University of Arizona, Tucson, AZ 85721, USA

\end{center}

\vspace{0.3cm}

\begin{abstract}
Weinberg's contributions to the power counting and 
derivation of few-nucleon forces in Chiral EFT
are briefly recalled. Subsequent improvements are reviewed,
concluding with the recent suggestion of a combinatorial enhancement.
\end{abstract}

\section{Introduction}

About thirty years ago, Steven Weinberg \cite{Weinberg:1990rz,Weinberg:1991um}
set in motion a new nuclear physics \cite{Hammer:2019poc}
based on the framework of effective field theory (EFTs), 
which he had formulated earlier \cite{Weinberg:1978kz}. 
An EFT includes all interactions allowed by the symmetries that are 
supported by 
the degrees of freedom relevant to the energies of interest. 
These interactions involve arbitrary numbers of 
derivatives and fields. One of the advantages over more phenomenological 
approaches, which attracted immediate attention in the nuclear community, 
is that few-body forces can be constructed consistently with the 
two-body force. 
By the time EFT came into the scene, various excellent phenomenological
parametrizations of the two-nucleon force existed
which failed to describe the three- and four-nucleon systems better 
than within about 20\%. Guessing the form of few-nucleon forces have proven
to be nearly impossible without EFT. 

Yet, the relative importance of few-body forces is not well established
in the Chiral EFT \cite{Hammer:2019poc} employed by Weinberg, 
which includes pions and nucleons.
A crucial ingredient of any EFT is the power counting that orders interactions 
according to the magnitudes of their contributions to observables,
but it requires assumptions which are rarely tested. 
Power counting is the rationale to neglect all but
a few interactions at each order,
thus enabling an {\it a priori} estimate of errors.

I arrived in Austin from S\~ao Paulo with an excellent background in physics,
at a time which allowed me to participate in the formulation of nuclear
EFTs.
I have recently related the events surrounding the early 
developments \cite{vanKolck:2021rqu}. 
In this brief report, I focus on the evolution of the ideas for
power counting few-body forces, starting with Weinberg's work, 
continuing with various 
subsequent improvements, and ending with the recent suggestion
of an environmental dependence on the number of nucleons.

\section{The problem}

QCD is the underlying theory of nuclear physics.
It is characterized by the nonperturbative scale
$M_{\rm QCD}\sim 1$ GeV, which is reflected in the masses of
most hadrons, including the nucleon's 
$m_N= {\cal O}(M_{\rm QCD})\simeq 940$ MeV. 
QCD has an approximate chiral symmetry,
whose spontaneous breaking generates pions of mass $m_\pi \simeq 140$ MeV
and interactions proportional to the inverse of the pion decay constant
$f_\pi={\cal O}(M_{\rm QCD}/4\pi)\simeq 92$ MeV.

We are interested here in systems of $A$ nucleons 
with typical momentum 
$Q\sim m_\pi \ll M_{\rm QCD}$. 
This is the domain of Chiral EFT \cite{Hammer:2019poc}, where nucleons
couple to pions 
according to the constraints of chiral symmetry.
If we integrate out all heavy mesons and baryon excitations,
observables can be calculated in an expansion in powers of $Q/M_{\rm QCD}$. 
For $A=0,1$, this theory reduces to Chiral Perturbation Theory (ChPT).
For perturbative amplitudes, the assumption of naturalness 
\cite{tHooft:1979rat,Veltman:1980mj}
---according to which the magnitude of
short-range interactions is set by their
bare parameters with the regulator cutoff replaced by $M_{\rm QCD}$---
gives rise to the so-called naive dimensional analysis (NDA) 
\cite{Manohar:1983md},
if one estimates that each loop contributes a factor of $(4\pi)^{-2}$.
This factor combines with factors of $f_\pi$ from the 
pion interactions to ensure a suppression
of $(Q/M_{\rm QCD})^2$ for each loop \cite{Weinberg:1978kz}.
Thus amplitudes are indeed perturbative, consistently with
the use of NDA in the first place.

For $A\ge 2$,
Weinberg \cite{Weinberg:1990rz,Weinberg:1991um} identified 
in $A$-nucleon reducible diagrams an infrared
enhancement by a relative factor of $m_N/Q$.
This was the first sign that nuclear amplitudes have a very 
different power counting than those for $A=0,1$, as the 
enhancement comes from nucleon recoil, a subleading effect in
ChPT.
Weinberg then defined the potential as the sum of irreducible subdiagrams,
which are not infrared enhanced.
The potential contains $a$-body components with $a=2, \ldots, A$.
An $a$-body force cannot be reduced, within the resolution of the EFT,
to an iteration of fewer-body forces; it is a force that disappears when 
any nucleon is removed. Even if the underlying theory were 
fundamental and its interactions of two-body character, the finite resolution of
the EFT would require the existence of few-body forces. Few-nucleon forces
are not forbidden by any symmetry, and therefore {\it must} appear at some order
in the EFT expansion of amplitudes. 
The question is, which order?

In his first paper \cite{Weinberg:1990rz}, Weinberg did not realize there
was a price to pay for building higher-body forces and ended
with the suggestion that few-body forces could be important.
He was quick to rectify
this oversight in his second paper \cite{Weinberg:1991um},
which announces the correction already in the abstract. 
The exchange of a single pion between two nucleons
brings to a diagram a factor of at most 
$4\pi/m_Nf_\pi$.
Weinberg assumed that, since the potential is 
free of infrared enhancements,
multi-nucleon contact interactions are also given by NDA, 
starting with the two-nucleon
contact (containing four nucleon fields in the Lagrangian) of size 
$4\pi/m_Nf_\pi$, too. 
These two-nucleon interactions are leading order (LO).
When one adds a nucleon to the force, one changes the number of loops in
the $A$-nucleon amplitude.
Assuming the same factor of $(4\pi)^{-2}$ as for ChPT loops,
Weinberg arrived at a cost of $(Q/M_{\rm QCD})^2$ for each additional
nucleon in the force. 

At that point we thought the first three-nucleon force appeared at relative 
${\cal O}(Q^2/M_{\rm QCD}^2)$ from diagrams 
where two nucleons interacted while there was already 
a pion ``in the air'' emitted by a third nucleon.
Prompted by a remark by James Friar, we realized these diagrams cancel against
the energy dependence in the one-pion-exchange 
two-nucleon force. However, this cancellation would only go through
if there was an error 
in expressions for pion-in-the-air diagrams
in Refs. \cite{Weinberg:1990rz,Weinberg:1991um}. 
When I pointed this out 
to Weinberg he quickly agreed, an example of his utmost intellectual
honesty that did wonders for
my self-esteem. The correct expression was published shortly
afterwards \cite{Ordonez:1992xp} and
details of the cancellation were given in Refs. 
\cite{Weinberg:1992yk,vanKolck:1994yi}.
As a consequence of this cancellation, 
the leading three-body force would come from interactions which are
themselves suppressed by 
one power of $Q/M_{\rm QCD}$.
That is, the first three-body force would appear at relative
${\cal O}(Q^3/M_{\rm QCD}^3)$, with four-body forces at 
${\cal O}(Q^4/M_{\rm QCD}^4)$ 
and so on. 

The leading components of the three-nucleon potential 
according to this power counting were derived in Refs. 
\cite{Ordonez:1992xp,vanKolck:1994yi,Epelbaum:2002vt}.
Sometimes referred to as the Texas potential, it has two-pion, 
pion/short-range, and purely short-range components.
The two-pion component is intimately related
to pion-nucleon scattering and 
carries the imprints of chiral symmetry. It slightly
corrects \cite{Friar:1998zt} the Tucson-Melbourne force \cite{Coon:1978gr}
to a form closer to the Brazil force \cite{Coelho:1983woa,Robilotta:1986nv}.
The pion/short-range component, in turn, is related to
$p$-wave pion production in nucleon-nucleon collisions \cite{Hanhart:2000gp},
while the purely short-range component is intrinsically a three-body
feature.
The shorter-range components have non-negligible effects on the 
three-body system \cite{Huber:1999bi,Epelbaum:2002vt} and beyond.
They have become very popular thanks to several successes,
such as an improved description of light nuclei \cite{Navratil:2007we}.

Unfortunately, there have not been extensive checks that these
order assignments are supported by data. It is remarkable
that many nuclear properties,
such as those of nuclear matter \cite{Drischler:2017wtt,Sammarruca:2018bqh},
are only described well with
chiral potentials based on Weinberg's power counting when
three-body forces are included. 
Most papers do not even report LO results.
In fact, nuclei beyond $A=4$ are not stable at LO \cite{Yang:2020pgi}.
Some of these problems are discussed in Ref. \cite{Tews:2020hgp}.

\section{``... and then we learn something''}

One of Weinberg's favorite remarks was that a theorist should insist 
on consistency with assumptions made, until evidence prompts 
their reevaluation
``and then we learn something''.
Sadly, Weinberg's papers have been accepted 
like a gospel by most nuclear physicists,
despite consistency issues that surfaced over time
which I address in the following.

\subsection{Role of the Delta}

The first issue is the role of the Delta isobar, whose mass
is only $\Delta \equiv m_\Delta-m_N\sim 300$ MeV above the nucleon's.
If one does not include an explicit degree of freedom for the Delta 
in Chiral EFT, its effects are subsummed into
contact interactions suppressed by powers
of $\Delta^{-1}$ \cite{Ordonez:1993tn,vanKolck:1994yi,Ordonez:1995rz}
instead of $M_{\rm QCD}^{-1}$.
Convergence is restricted. 

There is really no reason not to include an explicit Delta field.
When this is done, the leading three-nucleon force comes at
relative ${\cal O}(Q^2/M_{\rm QCD}^2)$ \cite{vanKolck:1994yi} 
in Weinberg's power counting.
In the form of the Fujita-Miyazawa force \cite{Fujita:1957zz},
it is the dominant component of the force.
One way \cite{Pandharipande:2005sx} 
to see the importance of the Delta is
to consider the relation between 
the two-pion component of the three-nucleon force 
and pion-nucleon scattering: 
one needs to extrapolate in energy by at least 
$m_\pi$ which leads to errors no smaller than ${\cal O}(m_\pi^2/\Delta^2)$
when the Delta is integrated out.
In contrast, with an explicit Delta one can extend the ChPT power counting 
to describe
pion-nucleon scattering through the Delta peak \cite{Long:2009wq}
and firmly determine pion-nucleon couplings.
Of course, the same argument holds for 
the two-pion components of two- and higher-body forces,
which should be constructed consistently \cite{Ordonez:1993tn,Ordonez:1995rz}.

\subsection{Loop factors}

The second shortcoming of Weinberg's power counting
is the estimate of the powers of $(4\pi)^{-1}$.
In the simpler Pionless EFT containing only 
nucleons \cite{Hammer:2019poc}, one can see explicitly that 
reducible loops have an additional enhancement of $4\pi$
relative to loops in ChPT. It is the combination of this enhancement
with the infrared enhancement of Weinberg's
that justifies \cite{Bedaque:2002mn} iterating the LO potential: 
a two-nucleon
reducible loop contributes an $m_NQ/4\pi$ that compensates the
additional $4\pi/m_Nf_\pi$ from the potential, leading at LO to
a series that needs resummation for $Q\sim f_\pi$ ---incidentally,
this generates naturally binding energies per nucleon
$B_A/A \sim 10$ MeV, as typically observed.
Counting $4\pi$s {\it \`a la} Weinberg will simply not do.

Taking into account the proper factor of $(4\pi)^{-1}$ for 
reducible loops,
Friar \cite{Friar:1996zw} arrived at an improved
power counting where few-nucleon forces are enhanced with respect
to Weinberg's. For more details, see Ref. \cite{vanKolck:2020llt}.
With an explicit Delta and Friar's counting the three-body force first
appears at next-to-leading order (NLO), that is, a relative
${\cal O}(Q/M_{\rm QCD})$ with respect to the LO two-body force.
Unfortunately, Friar's work is usually ignored by the nuclear 
community.

\subsection{NDA failure}

The third concern is the assumption of NDA. 
It is now well known that Weinberg's power counting is not consistent 
with the renormalization group (RG) at the two-body level 
\cite{Kaplan:1996xu,Nogga:2005hy,PavonValderrama:2005uj}.
The LO two-body potential in Chiral EFT is singular and its renormalization
requires more contact interactions than supplied by NDA \cite{vanKolck:2020llt}.
In hindsight, this might not be entirely surprising, as NDA is based
on perturbative renormalization. Continuing to assume naturalness,
but now in the appropriate nonperturbative context, 
leads to departures from NDA \cite{vanKolck:2020plz}.

One could then reasonably expect that NDA might breakdown
also in the many-body sector.
However, there is 
no RG evidence that either Weinberg's or Friar's countings fail 
for more-body forces once the two-nucleon amplitude is renormalized 
at LO and NLO
\cite{Nogga:2005hy,Song:2016ale,Yang:2020pgi}. 
This is in stark contrast with Pionless EFT where the RG demands a
three-body force at LO \cite{Bedaque:1998kg,Bedaque:1998km,Bedaque:1999ve}.
Based on continuity with Pionless EFT,
Kievsky and collaborators \cite{Kievsky:2016kzb}
suggested that three-nucleon forces should be included at LO
also in Chiral EFT. 
While there is an improvement in the description of data,
a power-counting rationale is missing. 

\section{A solution?}

The description of $A=3,4$ nuclei  
in properly renormalized
Deltaless EFT up to (and including) NLO ---that is, before three-nucleon
forces enter according to either Weinberg or Friar---
is actually very good \cite{Yang:2020pgi}. 
Thus in light nuclei few-body forces do not seem to be necessary at LO
in Chiral EFT, consistently with a lack of RG enhancement.
However, just as for potentials in Weinberg's power counting, 
larger nuclei are not stable at LO \cite{Yang:2020pgi}. 
While one cannot exclude that stability will emerge
at higher orders, which should be perturbative, 
instability could be a clue for the growing importance of
three-nucleon forces as $A$ increases. 

This led Jerry Yang and collaborators \cite{Yang:2021vxa} 
to propose
that the ordering of few-nucleon forces depends on the number of nucleons
present. 
It is not impossible that the power counting needs to be modified for
$A\gg 1$, as we then have an additional, large dimensionless
factor.
The basic idea is very simple: for $2<a< A/2$, there are ${}_A C_a/{}_A C_2$
more ways to construct 
an $a$-body than a two-body interaction, where 
${}_AC_a= A!/a!(A-a)!$ is the binomial coefficient. 
Of course, one needs to account as well for a suppression
by powers of $Q/M_{\rm QCD}$, and the question arises of the dependence of
the typical bound-state momentum $Q$ on $A$.
There is no obvious answer, except for $A=2$
where the position of the pole in the imaginary axis
of the complex-momentum plane
is $(2m_N B_A/A)^{1/2}$.
This is the same as one would naively guess by assuming each nucleon 
contributes $Q^2/m_N$ to $B_A$. With this assumption, 
the fact that $B_A/A$ is essentially
constant for $A\ge 4$ would lead to a constant $Q$.
With Friar's counting in Deltaless Chiral EFT,
the suppression is $(Q/M_{\rm QCD})^2$.
If $Q\sim 3f_\pi$ for nuclear matter, this suppression can be alternatively
written as $\rho_0/f_\pi^2 M_{\rm QCD}$, where $\rho_0\simeq 0.16$ fm$^{-3}$
is the saturation density.
With these very rough estimates, one expects 
three-nucleon forces to become comparable to two-nucleon forces
for $A\sim 20$, quickly followed by four-body forces at $A\sim 25$. 

While these critical values of $A$ cannot be taken very seriously, they suggest 
there might be a range of nuclei for $A>4$ where three-body forces
should be included at LO, despite the fact that they are subleading
(and thus perturbative) for $A\le 4$. It is encouraging that then $^{16}$O
and even $^{40}$Ca become stable \cite{Yang:2020pgi}. 
However, for the latter the single-particle states indicate a disfavored
deformation, which could be a consequence of the inappropriate
neglect of four-body forces at such large $A$. 
Since on account of the exclusion principle
five- and more-nucleon forces have additional $Q/M_{\rm QCD}$ suppression,
it is possible that they never become important.

Even if the combinatorial factor is not the root cause of 
an enhancement of three-body forces, it could be that resumming
these forces into LO for $A> 4$ is justified as an ``improved action'' 
in the sense of 
of lattice QCD: an interaction that is introduced to accelerate
convergence ---in this case, to obtain stability already at LO 
without breaking RG invariance and to enable 
a perturbative treatment of corrections.

\section{Conclusion}

Despite the importance of few-nucleon forces for a consistent description
of nuclei and the many years of development in Chiral EFT,
the order at which they should first be included remains to be conclusively
established.
We hope that a better understanding of the leading order ---which 
should give the correct physics within the error of the EFT
expansion but is usually avoided by potential modelers--- will soon emerge 
thanks to improved ``{\it ab initio}'' methods \cite{Hergert:2020bxy}
for the solution the many-body Schr\"odinger equation.

\vspace{1cm}
\noindent
{\bf Acknowledgements.}
I thank Manuel Malheiro and Nilberto Medina 
for the invitation to speak at 
XLIV RTFNB.
It is always a pleasure to interact with the Brazilian nuclear community, 
even if remotely.
This work was supported in part
by the U.S. Department of Energy, Office of Science, Office of Nuclear Physics,
under award number DE-FG02-04ER41338.

\end{document}